\documentclass[10pt,twocolumn]{article}
\usepackage[letterpaper,margin=0.7in,columnsep=0.25in]{geometry}
\usepackage[T1]{fontenc}
\usepackage{lmodern}
\usepackage{amsmath,amssymb,bm,graphicx,xcolor,url}
\usepackage[labelfont=bf,labelsep=period,font=small]{caption}
\usepackage[numbers,sort&compress,round]{natbib}
\usepackage{authblk}
\usepackage{pdfpages}
\usepackage[colorlinks=true,allcolors=blue]{hyperref}

\newcommand{\sref}[1]{\textcolor{blue}{#1}}
\renewcommand{\eqref}[1]{\textup{(\ref{#1})}}
\graphicspath{{./}{Figures/}}
\newcommand{\Parasplit}{ }
\makeatletter
\newcommand{\matmethods}[1]{\gdef\@matmethods{#1}}
\newcommand{\showmatmethods}{\section*{Materials and Methods}\@matmethods}
\newcommand{\dataavail}[1]{\subsection*{Data, Materials, and Software Availability}#1}
\newcommand{\acknow}[1]{\gdef\@acknow{#1}}
\newcommand{\showacknow}{\subsection*{Acknowledgments}\@acknow}
\renewcommand{\@biblabel}[1]{#1.}
\makeatother

\title{\bfseries Time--temperature--composition equivalence of the glass transition in digitally mixed 3D-printed polymers}
\author[1]{Beijun Shen}
\author[2]{Sanat K. Kumar}
\author[1,*]{Mary C. Boyce}
\affil[1]{Department of Mechanical Engineering, Columbia University, New York, NY 10027}
\affil[2]{Department of Chemical Engineering, Columbia University, New York, NY 10027}
\date{}

\begin{document}
\twocolumn[{\begin{@twocolumnfalse}
\maketitle
\begin{abstract}
For more than seventy years, time--temperature equivalence (TTE) has enabled a few measurements to describe the linear viscoelasticity of polymers over many decades of frequency. Multimaterial 3D printing now combines elastomeric and glassy resins voxel by voxel into digitally mixed polymers. Composition shifts the glass transition along the frequency axis, as temperature does, but also broadens it and increases the rubbery plateau modulus. A translation, the only operation of TTE, cannot capture these changes. Here we establish a \textit{time--temperature--composition equivalence} (TTCE) from dynamic mechanical measurements of five compositions of a printed family. A shift of the transition along the reduced-frequency axis, a stretch of its breadth and a normalization between its plateaus collapse the storage modulus onto one master curve. The collapse holds at every measured temperature, frequency and composition, through a single reduced variable. The position and breadth of the transition change linearly with the volume fraction of the glassy resin. The phase angle and the relaxation spectrum superpose with the same parameters, so the family shares one distribution of relaxation times. The resulting closed form gives the storage and loss moduli continuously in composition, temperature and frequency, and quantitatively predicts two further mixtures withheld from the calibration. The same shift, stretch and normalization also collapse published temperature sweeps of a patterned composite family of different resins, pointing to broader applicability. Composition thus joins temperature and frequency as a third variable of one equivalence, and the closed form can guide the design of graded and architected printed structures.
\end{abstract}
\vspace{1.5em}
\end{@twocolumnfalse}}]
{\renewcommand{\thefootnote}{}\footnotetext{\textsuperscript{*}Corresponding author. E-mail: mb3814@columbia.edu}}

Multimaterial 3D printing deposits different polymers voxel by voxel, so that composition becomes a design variable chosen point by point within a part~\cite{TrubyLewis2016,SkylarScott2019}. Photopolymer jetting (PolyJet) does so by depositing droplets of liquid photopolymers and curing each layer with ultraviolet light~\cite{Mueller2017,Zorzetto2020}. Jetting an elastomeric resin (Agilus) and a glassy resin (Vero) in prescribed ratios produces a family of \textit{digitally mixed} polymers. At room temperature the family ranges from a soft viscoelastic elastomer to a stiff glassy polymer. We take \textit{elastomeric} to denote a compliant, time-dependent response above the glass transition, and \textit{glassy} to denote a stiff response below the glass transition. Composition and its spatial arrangement underlie architected and energy-absorbing composites~\cite{Wang2011,Mora2022,Kim2025}, functionally graded structures~\cite{FloresItuarte2019,Kuang2019,Ahn2024}, shape-memory composites~\cite{Yuan2020,Yuan2021}, and viscoelastic metamaterials and wearable devices~\cite{KarimiMahabadi2026,Wirth2025}. Across the family, the glass transition moves from below to well above room temperature, so stiffness and damping change by orders of magnitude within ordinary service conditions. The response of a printed part thus depends on three variables: the composition assigned at printing, and the temperature and frequency imposed in service. Because the composition can take any value between those of the two base resins, measuring stiffness and damping over temperature and frequency for each composition\Parasplit separately is impractical. Here we show that a single equivalence describes the linear viscoelasticity of the entire digitally mixed family. This \textit{time--temperature--composition equivalence} (TTCE) contains the classical time--temperature equivalence (TTE) as a special case and adds composition as a third variable.

Temperature and frequency are already linked by classical TTE. A polymer is rubbery when its chain segments can sample space within the time scale of loading, and glassy when intermolecular interactions hinder that sampling. Decreasing the temperature lengthens the time needed for segmental relaxation, and increasing the frequency shortens the time available. Cooling and an increase in frequency therefore both carry the polymer toward the glassy state, and a change in temperature can be offset by a change in frequency. Each isotherm, a frequency sweep at one temperature, is shifted along the logarithmic frequency axis by a factor $a_T$. The storage modulus and loss factor then collapse onto one master curve~\cite{WLF1955,Ferry1980}, and a few measurements describe the response over many decades of frequency. Here and throughout, a \textit{decade} denotes a factor of ten on a logarithmic scale of the quantity named. TTE thus operates by translation alone. It moves the transition along the frequency axis and leaves its shape, including its breadth, unchanged. TTE rests on the assumption that segmental relaxation processes all share the same scaling with temperature and hence the horizontal shift.

TTE has since been generalized to other variables. Shifts along the time or frequency axis extend it to polymer concentration~\cite{LiWinter2023}, pressure~\cite{Bair2014}, solvent and moisture content~\cite{SuarezMartinez2019,Liu2025}, degree of cure~\cite{AdolfMartin1990,Phansalkar2026} and cross-link density~\cite{Plazek1966}. Generalized superposition over the length and concentration of F-actin filaments also rescales time~\cite{Semmrich2007}. In time--connectivity superposition, the shape of the relaxation spectrum is set at the gel point and kept as the gel ages~\cite{Keshavarz2021}. In nanofilled polymers, time--temperature--concentration superposition builds master curves over filler content~\cite{Song2015,Liu2009}, with the filler adding to the modulus while the segmental relaxation of the matrix stays in place~\cite{Baeza2016,Mujtaba2014,Robertson2008}. Time--composition superposition has been reported for block-copolymer solutions~\cite{Krishnan2012} and, through vertical shifts alone, for immiscible vitrimer blends~\cite{Torres2026}. When composition changes the shape of the transition, the relaxation processes clearly depend on composition and therefore a single horizontal translation no longer suffices. In plasticized polystyrene, the reduction over concentration requires different shift factors in the transition and terminal zones~\cite{Plazek1979}. In miscible blends, the Fox, Gordon--Taylor and Couchman--Karasz relations give $T_g$ as a function of composition~\cite{Fox1956,GordonTaylor1952,CouchmanKarasz1978} but not the breadth of the transition. TTE holds in such blends when the component $T_g$ are close~\cite{Pathak1998} but fails when they are far apart, because the local composition then broadens the transition~\cite{Colby1989,Pathak1999,LodgeMcLeish2000}.

The digitally mixed family is a well-defined system in which to probe whether composition may act in a manner analogous to temperature and frequency. The two acrylate resins co-cure and interdiffuse at the voxel scale into one cross-linked solid~\cite{Yuan2021,LeeCho2026}. Each mixture shows a single glass transition, and its $T_g$ increases with the Vero fraction~\cite{Ho2024}. The response is nonetheless characterized mixture by mixture~\cite{Slesarenko2018,Daneshdoost2024}, and design methods draw on databases of measured responses~\cite{KarimiMahabadi2026}. The closest study built master curves for seven compositions and predicted untested ones from a percolation fit, without making composition a variable of the superposition~\cite{Wirth2025}. In our recent work, we built a composition-dependent finite-strain constitutive model of the family~\cite{ShenBoyceJMPS} and data-driven models of its Agilus-rich mixtures~\cite{GarciaAvila2026}. The dynamic measurements of ref.~\cite{ShenBoyceJMPS} showed that the glass transition moves to higher temperature with the Vero fraction and with frequency alike. Composition, however, also broadens the transition and increases the rubbery plateau modulus. Whether a single equivalence can capture the position, breadth and plateaus of an entire printed family has remained open.

\begin{figure*}[tbp]
\centering
\includegraphics[width=\textwidth]{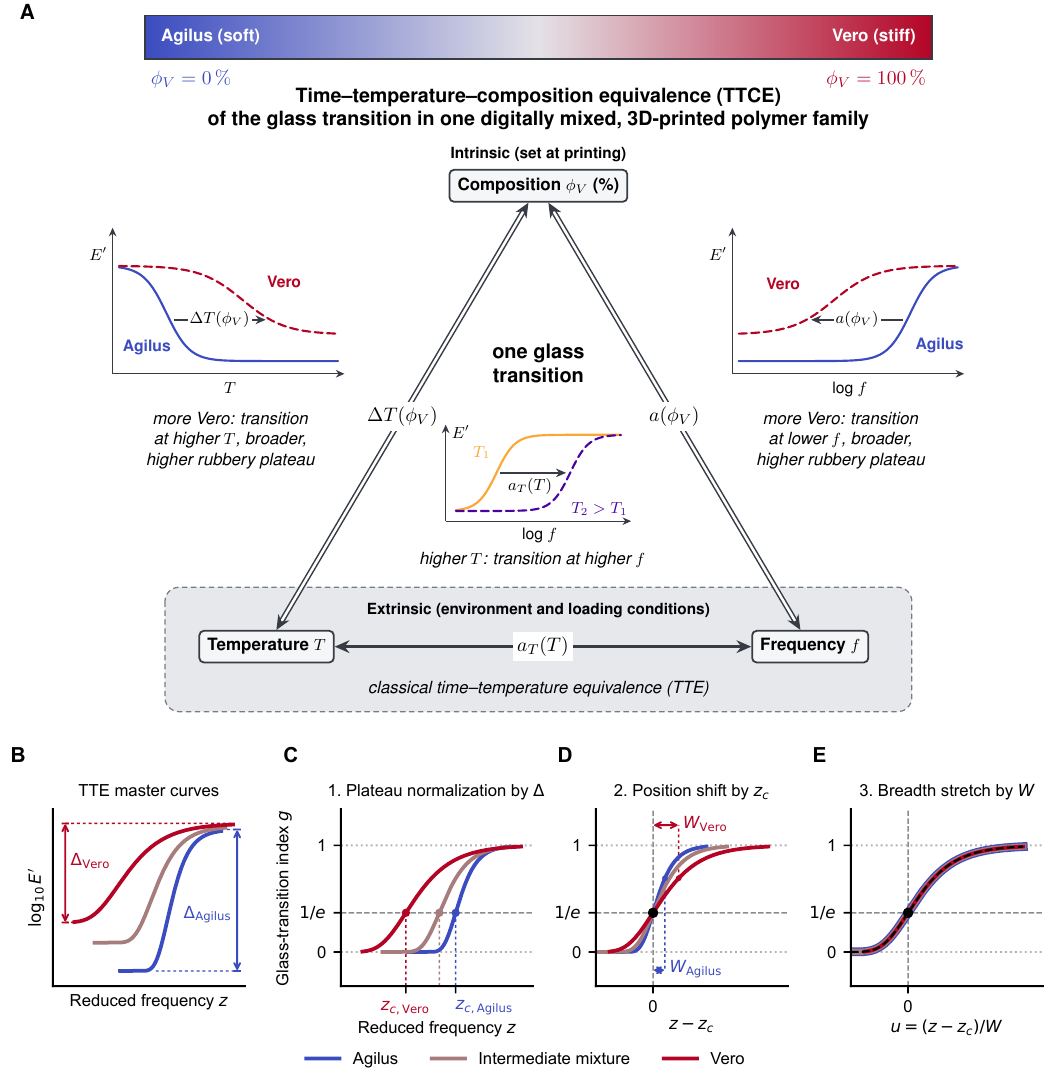}
\caption{\textbf{Time--temperature--composition equivalence of the glass transition in one digitally mixed, 3D-printed polymer family.} (A) Temperature $T$ and frequency $f$ are loading conditions, linked by the shift factor $a_T(T)$ of time--temperature equivalence (TTE). The Vero volume fraction $\phi_V$ is fixed at printing and moves the glass transition by $a(\phi_V)$ along the frequency axis and by $\Delta T(\phi_V)$ along the temperature axis. The sketches show the effect of each variable on the storage modulus $E'$. Temperature translates the transition. Composition translates it, stretches it and increases its rubbery plateau. (B) Sketched TTE master curves of Agilus, an intermediate mixture and Vero. Arrows mark $\Delta$. (C--E) The three operations of the TTCE. (C) Plateau normalization by $\Delta$ (Eq.~\eqref{eq:form}) gives the glass-transition index $g$. Dots mark the steepest point of each curve, $g=1/e$ at $z=z_c$. (D) The position shift moves these points to $z-z_c=0$. Arrows mark the breadth $W$. (E) The breadth stretch divides $z-z_c$ by $W$ and superposes all curves on $g(u)$ of Eq.~\eqref{eq:master} (dashed).}
\label{fig:concept}
\end{figure*}

We introduce the TTCE and establish it experimentally for the glass transition and the linear viscoelastic response of the family (Fig.~\ref{fig:concept}). Temperature and frequency are extrinsic loading conditions, whereas composition, given by the Vero volume fraction $\phi_V$, is intrinsic to the material and fixed at printing. Composition translates the glass transition along the frequency axis, as temperature does. The position and the breadth of the transition are both found to change linearly with the Vero fraction. A translation, a stretch and a normalization by the plateaus are found to collapse the storage modulus of five compositions onto one master curve. The collapse holds at every measured temperature and frequency, through a single reduced variable. The phase angle and the relaxation spectrum superpose with the same parameters, so the family shares one distribution of relaxation times, albeit shifted and stretched in time. A closed form then gives stiffness and damping continuously in composition, temperature and frequency. Five compositions, each measured at four frequencies, suffice to cover every composition between the two base resins, over more than twenty decades of frequency. The closed form also predicts the measured stiffness and damping of two further mixtures not used in its calibration. The TTCE thus turns a printed family from a set of separately measured materials into effectively a single material, with composition as a third variable of its viscoelasticity. The same three operations also collapse the published temperature sweeps of Yuan et al.~\cite{Yuan2021}. Their family combines a different pair of resins in a pattern of $300~\mu$m cells, a length scale much larger than the voxel-by-voxel mixing of the present family.

\section{Results}\label{sec:results}

\subsection{One glass transition and one time--temperature equivalence per composition}\label{sec:onefamily}\label{sec:tf}

\begin{figure*}[tbp]
\centering
\includegraphics[width=0.9\textwidth]{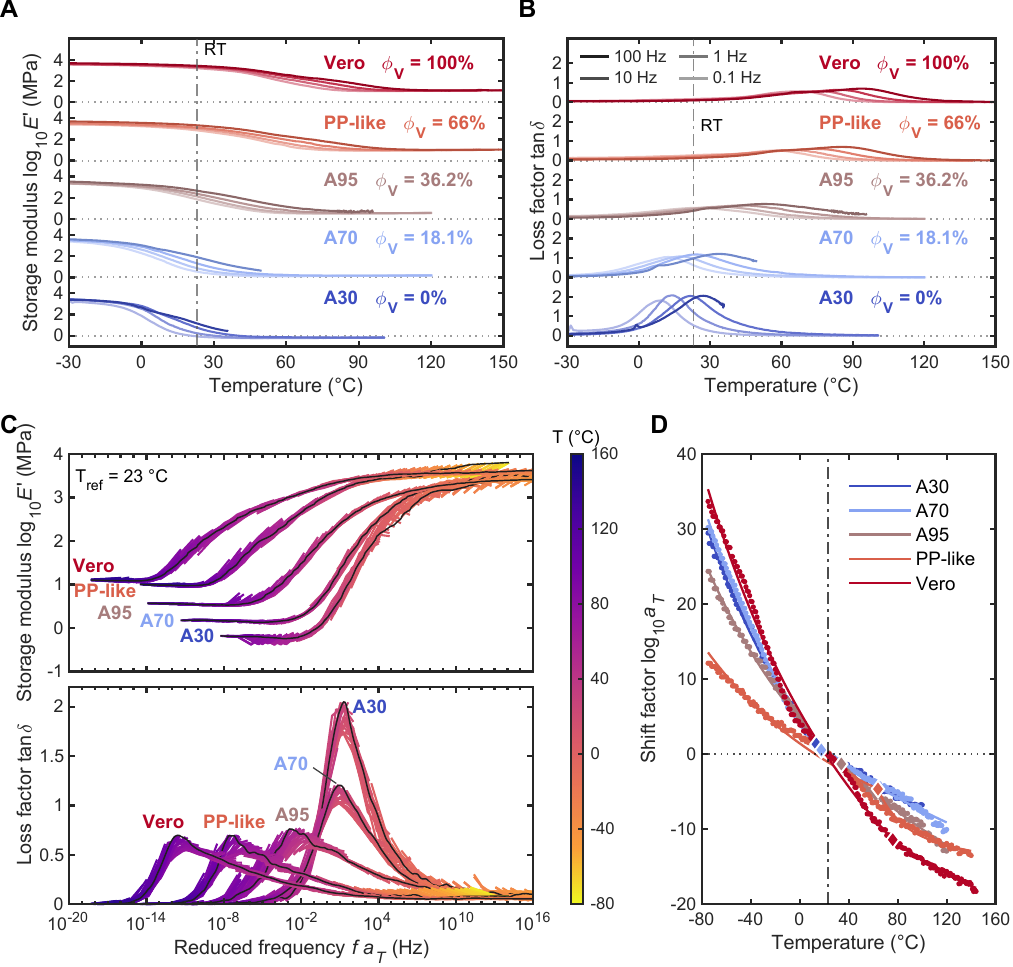}
\caption{\textbf{One glass transition and one time--temperature equivalence per composition.} (A) Storage modulus and (B) loss factor against temperature at $0.1$, $1$, $10$ and $100$~Hz. Compositions are offset vertically, each with its own axis. Color denotes composition, and darker lines denote higher frequency. (C) TTE master curves of the storage modulus (top) and loss factor (bottom), referenced to $23~^{\circ}$C (RT). Isotherms are colored by temperature, and black curves are the smoothed TTE master curves. Both quantities share one shift factor $a_T$ per temperature. (D) Measured shift factors (points) and piecewise Arrhenius and Williams--Landel--Ferry fits (lines, \textit{SI Appendix}, Table~\sref{S1}). Diamonds mark $T_g$. Dash-dotted lines in A, B and D mark RT, where $a_T=1$. The data of A--C were first reported in ref.~\cite{ShenBoyceJMPS}. \textit{SI Appendix}, Fig.~\sref{S2} shows each composition, with the loss modulus.}
\label{fig:dma}
\end{figure*}

Fig.~\ref{fig:dma} A and B show the storage modulus $E'$ and loss factor $\tan\delta$ of five compositions at four frequencies from $0.1$ to $100$~Hz. The compositions are A30, A70, A95, PP-like and Vero, with Vero volume fractions $\phi_V$ of $0$, $18.1$, $36.2$, $66$ and $100\%$ (Methods). The storage modulus of each composition decreases by two and a half to nearly four decades of modulus across its glass transition. The transition moves to higher temperature with an increase in Vero fraction and, within each composition, with an increase in frequency. Composition and frequency therefore move the glass transition in the same direction.

Each composition has one glass transition. Every sweep shows a single maximum in the loss factor. The weak shoulder below the main peak of neat Vero, reported in our earlier work~\cite{ShenBoyceJMPS}, does not resolve into a separate maximum. By contrast, 3D-printed composites of an elastomeric and a glassy resin patterned at a coarser scale show two distinct maxima, one near the $T_g$ of each resin~\cite{Yuan2020,LeeIJSS2026,Yuan2021}. The two maxima merge into one when the feature size is reduced to $300~\mu$m~\cite{Yuan2021}. The present resins are mixed voxel by voxel, far below this size, in layers $27~\mu$m thick (Methods). Each mixture is therefore found to behave as one polymer.

The two usual criteria for $T_g$ are the onset of the $E'$ downturn and the $\tan\delta$ peak. Both increase with Vero fraction and frequency, but by different amounts (\textit{SI Appendix}, Fig.~\sref{S1}). A single temperature therefore does not characterize a transition this broad, and the analysis below considers the entire curve.

TTE is then established for each composition. Shifting each isotherm along $\log_{10}f$ (Methods) forms the master curves of Fig.~\ref{fig:dma}C. They extend the three measured decades of frequency to more than twenty decades of reduced frequency (see \textit{SI Appendix}, Fig.~\sref{S2} for each composition). The shift factors span $26$ to $52$ decades of $a_T$ (Fig.~\ref{fig:dma}D). They follow an Arrhenius form below $T_g$ and the Williams--Landel--Ferry form above it (\textit{SI Appendix}, Table~\sref{S1}). The same shift factor also superposes the loss factor (Fig.~\ref{fig:dma}C, bottom). This superposition is the stricter test. A second relaxation process would distort the peak of the loss factor, as in blends whose two components keep distinct phases and glass transitions~\cite{Pathak1999}. TTE thus holds for all five compositions, as reported for PolyJet elastomers~\cite{Abayazid2020} and for stochastic mixtures of Agilus and Vero~\cite{Wirth2025}. Within each composition, temperature and frequency enter through the single reduced frequency $f\,a_T$.

The five master curves differ in position, breadth and plateaus. The increase of the storage modulus from the rubbery to the glassy plateau spans approximately $13$ decades of reduced frequency for A30 and $24$ for Vero. The wider span indicates a more heterogeneous molecular structure and a broader relaxation spectrum in the Vero-rich mixtures. The rubbery plateau increases with the Vero fraction, whereas the glassy plateau changes far less (Fig.~\ref{fig:dma}C, top). The peak of the loss-factor master curve decreases from $2.0$ in A30 to about $0.7$ in PP-like and Vero. Composition thus changes the shape and the plateaus of the transition as well as its position. A translation, which superposes temperature and frequency within each composition, cannot superpose the compositions. Section~\ref{sec:fphi} shows that a single equivalence does.

\subsection{Time--temperature--composition equivalence of the storage modulus of the family}\label{sec:fphi}

\begin{figure*}[tbp]
\centering
\includegraphics{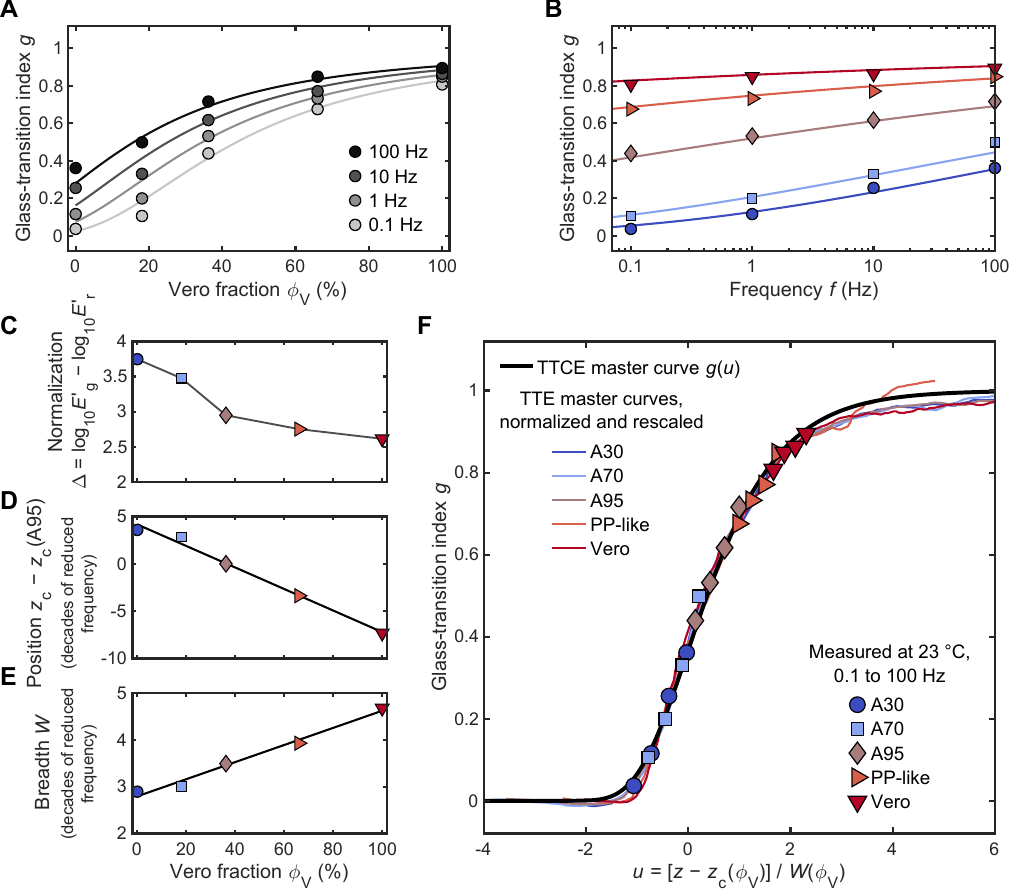}
\caption{\textbf{Time--temperature--composition equivalence of the storage modulus of the family.} (A and B) Glass-transition index $g$ at $23~^{\circ}$C against Vero fraction (A, one curve per frequency) and against frequency (B, one curve per composition). Points are measured. Curves are Eq.~\eqref{eq:master}, with $z_c$ and $W$ from the linear laws in A and from the fit to each composition in B. (C--E) Normalization $\Delta$, position $z_c$ (relative to A95, to center the axis) and breadth $W$ against Vero fraction (\textit{SI Appendix}, Table~\sref{S2}). The line in C joins the points, and the lines in D and E are Eqs.~\eqref{eq:shift} and~\eqref{eq:breadth}. (F) The five TTE master curves (thin lines) and the twenty points of A and B, normalized by Eq.~\eqref{eq:form} and plotted against $u$ of Eq.~\eqref{eq:u}. All collapse onto $g(u)$ (black).}
\label{fig:gd}
\end{figure*}
These differences are sketched in Fig.~\ref{fig:concept}B for Agilus, an intermediate mixture and Vero. We capture all three in one expression by writing the storage modulus of each composition as
\begin{equation}
\log_{10}E'=\log_{10}E'_r+\Delta\cdot g(u),\qquad \Delta=\log_{10}E'_g-\log_{10}E'_r,
\label{eq:form}
\end{equation}
where $E'_r$ and $E'_g$ are the rubbery and glassy plateau values of the master curve of that composition. The normalization $\Delta$ is the increase of $\log_{10}E'$ from rubbery to glassy (Fig.~\ref{fig:concept}B). The glass-transition index $g$ is thus the storage modulus normalized on a logarithmic scale, $0$ on the rubbery plateau and $1$ on the glassy plateau. This plateau normalization, the first of three operations, removes the difference in plateaus (Fig.~\ref{fig:concept}C). The index $g$ depends on composition, temperature and frequency through one reduced variable $u$,
\begin{equation}
u=\frac{z-z_c(\phi_V)}{W(\phi_V)},\qquad z=\log_{10}f+\log_{10}a_T(T;\phi_V).
\label{eq:u}
\end{equation}
Here $z$ is the reduced frequency of TTE, referenced to $23~^{\circ}$C (Section~\ref{sec:tf} and Fig.~\ref{fig:dma}D). The position $z_c$ is the reduced frequency of the steepest point of the transition of a composition. The breadth $W$ sets the scale of the transition along $z$. Both $z_c$ and $W$ are in decades of reduced frequency and depend on composition, as discussed below. Against $u$, the index of every composition follows one master curve,
\begin{equation}
g(u)=\exp\!\left[-\exp(-u)\right].
\label{eq:master}
\end{equation}
The form of Eq.~\eqref{eq:master} is chosen because the measured transitions are asymmetric. In Fig.~\ref{fig:dma} A and C, the storage modulus turns sharply at the rubbery plateau but broadly and gradually at the glassy plateau, most visibly for the Vero-rich compositions. Equation~\eqref{eq:master} has the same asymmetry: $g(u)$ increases steeply from $0$ and approaches $1$ gradually. In reduced frequency for each composition, Eq.~\eqref{eq:master} is a stretched exponential, $g=\exp[-(f_c/f_r)^{\beta}]$ with $f_r=f\,a_T$, $f_c=10^{z_c}$ and $\beta=1/(W\ln 10)$. Equation~\eqref{eq:master} thus gives a composition dependence to the established Kohlrausch--Williams--Watts form of relaxation in polymers~\cite{Kohlrausch1854,Williams1970}. The breadth $W$ is inversely proportional to the stretching exponent $\beta$, and $\beta$ decreases as the distribution of relaxation times widens~\cite{Lindsey1980}. Here $\beta$ is $0.15$ for A30 and $0.09$ for Vero (\textit{SI Appendix}, Table~\sref{S2}).

The second operation, the position shift, aligns the normalized curves at their steepest points. By Eq.~\eqref{eq:u}, the steepest point of each curve, at $z=z_c$, lies at $u=0$, where Eq.~\eqref{eq:master} gives $g=1/e\approx0.37$ (dots in Fig.~\ref{fig:concept}C). Subtracting $z_c$ from $z$ translates each normalized curve until its steepest point lies at $z-z_c=0$ (Fig.~\ref{fig:concept}D). The shifted curves still differ in breadth. The breadth $W$ of each curve is the interval of reduced frequency from its steepest point ($u=0$) to $u=1$ (arrows in Fig.~\ref{fig:concept}D). The breadth stretch, the third operation, divides $z-z_c$ by $W$ and so rescales the reduced-frequency axis of each composition by its own breadth. All curves then superpose on a single TTCE master curve $g(u)$ (Fig.~\ref{fig:concept}E).

Fig.~\ref{fig:gd} tests the equivalence. Fig.~\ref{fig:gd} A and B show twenty values of $g$ measured at $23~^{\circ}$C, one per composition and frequency. They are plotted against the Vero fraction at each frequency (A) and against frequency for each composition (B). Each value is the measured storage modulus of one composition, normalized between the two plateaus of that composition by Eq.~\eqref{eq:form}. Increasing the Vero fraction and increasing the frequency carry a material along the same asymmetric S-shaped increase of $g$ from the rubbery to the glassy state. Although normalized, the response of each composition is still centered at its own frequency and has its own breadth. Bringing all five compositions onto one curve therefore requires the position shift and the breadth stretch of Fig.~\ref{fig:concept} D and E.

The plateaus are read from the two ends of each TTE master curve and set the normalization $\Delta$ (Fig.~\ref{fig:gd}C). Across the family the rubbery plateau $E'_r$ increases twentyfold, from $0.6$ to $12$~MPa, while the glassy plateau $E'_g$ stays between $3.2$ and $5.5$~GPa. The normalization $\Delta$ therefore decreases from $3.7$ to $2.6$ decades of modulus.

The position $z_c$ and breadth $W$ of each composition are found by fitting Eqs.~\eqref{eq:form} to~\eqref{eq:master} to the whole TTE master curve of that composition (Methods). Each master curve contains every measured isotherm. No composition serves as a reference, because each transition is located by its own steepest point $z_c$. The position and breadth are the only fitted quantities, and both are linear in the Vero fraction. The position (Fig.~\ref{fig:gd}D) follows
\begin{equation}
z_c(\phi_V)=2.67-11.35\,\phi_V.
\label{eq:shift}
\end{equation}
Composition therefore translates the transition by approximately $11$ decades of reduced frequency from A30 to Vero. The breadth (Fig.~\ref{fig:gd}E) increases from $2.9$ decades of reduced frequency in A30 to $4.7$ in Vero,
\begin{equation}
W(\phi_V)=2.80+1.82\,\phi_V.
\label{eq:breadth}
\end{equation}

Fig.~\ref{fig:gd}F, the central test, applies the three operations of Fig.~\ref{fig:concept} C--E to the measured TTE master curves of Fig.~\ref{fig:dma}C. These curves cover every measured temperature and frequency of all five compositions. All five master curves superpose on the single curve of Eq.~\eqref{eq:master}, to within $0.03$ in $g$ (root mean square, Methods). So do the twenty room-temperature values of Fig.~\ref{fig:gd} A and B.

In summary, Eqs.~\eqref{eq:form} to~\eqref{eq:breadth} provide the TTCE for the storage modulus. Temperature and frequency act only through $z$: they translate the transition of a composition along the frequency axis and leave its shape unchanged. Classical TTE (Section~\ref{sec:tf}) is therefore the translation-only special case of the TTCE. Composition translates the transition, stretches it and moves its plateaus. Under TTE, temperature multiplies the frequency by $a_T$. Under the TTCE, composition also raises $f\,a_T/f_c$ to the power $1/W$, since $10^{u}=(f\,a_T/f_c)^{1/W}$. The reduced variable $u$ locates a material within its glass transition. Any two combinations of composition, temperature and frequency that give the same $u$ have the same index $g$. They lie at the same point of the master curve of Eq.~\eqref{eq:master}. The next section shows that the same parameters superpose the loss and the relaxation spectrum as well.

\subsection{The equivalence extends to the loss and the relaxation spectrum}\label{sec:loss}

\begin{figure*}[tbp]
\centering
\includegraphics{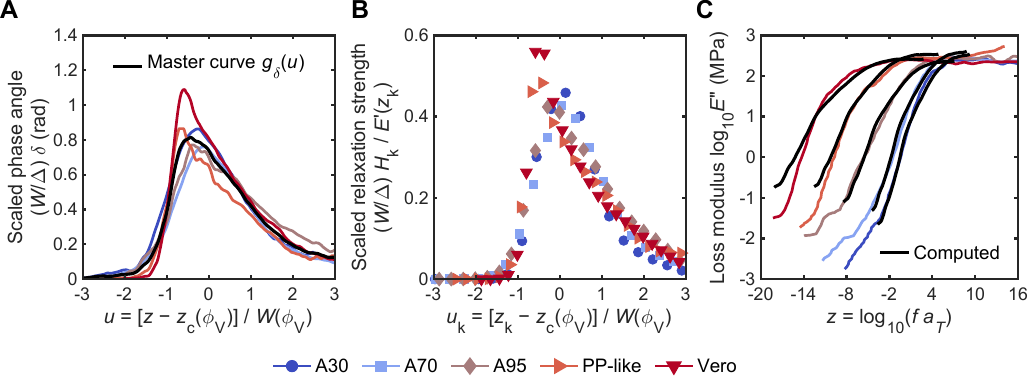}
\caption{\textbf{The equivalence extends to the loss and the relaxation spectrum.} All panels use the plateaus, $z_c$ and $W$ of each composition (\textit{SI Appendix}, Table~\sref{S2}), with no new parameter. (A) Phase angle $\delta=\arctan(\tan\delta)$ multiplied by $W/\Delta$, against $u$. The black curve, the mean of the five curves, is the master curve $g_\delta(u)$. (B) Relaxation strength $H_k$ of each Prony mode, divided by $E'(z_k)$ and by $\Delta/W$, against $u_k$. \textit{SI Appendix}, Fig.~\sref{S3}C shows the unscaled spectra. (C) Loss modulus against reduced frequency, referenced to $23~^{\circ}$C. Colored curves are the smoothed TTE master curves. Black curves are $E''=E'\tan\delta$ computed from Eqs.~\eqref{eq:form} to~\eqref{eq:master} and $g_\delta(u)$ across the transition.}
\label{fig:loss}
\end{figure*}

The data of Fig.~\ref{fig:dma}B show that the $\tan\delta$ peak shifts to higher temperature, broadens and decreases in height with increasing Vero fraction. Fig.~\ref{fig:loss} tests whether the position, breadth and plateau operations that provide the TTCE for the storage modulus also capture the loss response. The storage modulus and the loss are two views of one relaxation spectrum, and the loss is largest where the storage modulus increases most steeply with frequency~\cite{Ferry1980,SchwarzlStaverman1953}. Under the TTCE of Eqs.~\eqref{eq:form} and~\eqref{eq:u}, the storage modulus of every composition increases by $\Delta$ decades of modulus through the same shape in $u$. One unit of $u$ spans $W$ decades of reduced frequency. The steepness of the storage modulus, and with it the loss, therefore carries composition only through the ratio $\Delta/W$.

Fig.~\ref{fig:loss}A tests the equivalence on the phase angle $\delta=\arctan(\tan\delta)$, read from the smoothed TTE master curves of the loss factor (Fig.~\ref{fig:dma}C and \textit{SI Appendix}, Fig.~\sref{S2}). Note that the phase angle is used rather than the loss factor ($\tan\delta$) because it is bounded by $\pi/2$, whereas the loss factor diverges as $\delta$ approaches $\pi/2$ and exaggerates the tallest peaks. The TTE master curve of the phase angle of each composition is multiplied by $W/\Delta$ of that composition and plotted against $u$. All parameters come from the storage modulus (Fig.~\ref{fig:gd}), so no new parameter enters. As seen in Fig.~\ref{fig:loss}A, the five curves coincide in position and breadth, further demonstrating the TTCE. Their mean, shown in black, is the master curve of the phase angle, $g_\delta(u)$, the counterpart of $g(u)$ for the loss. The loss is a stricter test than the storage modulus, because it follows the slope of the storage modulus, and a slope magnifies small differences in shape. The peak heights of the five curves therefore differ somewhat. The threefold decrease of the peak loss factor across the family (Section~\ref{sec:tf}) reflects the smaller $\Delta$ and the larger $W$ of the Vero-rich compositions (Fig.~\ref{fig:gd} C and E).

Fig.~\ref{fig:loss}B tests the equivalence on the relaxation spectrum itself. A Prony series, with one relaxation mode per decade of relaxation time, was fitted to the TTE master curves of each composition (Methods and \textit{SI Appendix}, Fig.~\sref{S3}). Mode $k$ has relaxation time $\tau_k$, strength $E_k$ and reduced frequency $z_k=\log_{10}f_k$ at $23~^{\circ}$C, with $f_k=1/(2\pi\tau_k)$. Its relaxation strength $H_k=E_k/\ln 10$ is the strength per unit $\ln\tau$, and the set of $H_k$ is the discrete relaxation spectrum~\cite{Ferry1980,BaumgaertelWinter1989}. The spectrum at $23~^{\circ}$C covers every temperature, because temperature shifts all relaxation times of a composition by the same factor $a_T$. Fig.~\ref{fig:loss}B plots each mode at $u_k=[z_k-z_c(\phi_V)]/W(\phi_V)$, the reduced variable of Eq.~\eqref{eq:u}. The ordinate is the fraction of the storage modulus that relaxes per unit $\ln\tau$, $H_k/E'(z_k)$, divided by $\Delta/W$ as for the phase angle. The five scaled spectra coincide in position and breadth, again demonstrating the TTCE.

The scaled spectra have the same form as $g_\delta(u)$ in Fig.~\ref{fig:loss}A. Each increases steeply on the rubbery side, peaks at about the same $u$ and decreases slowly on the glassy side. The five compositions thus share one distribution of relaxation times in terms of the reduced variable $u$. Composition moves this distribution along the reduced-frequency axis and stretches it, by the same position shift and breadth stretch that superpose the storage modulus in Fig.~\ref{fig:gd}. Consistent with this stretch, the transition of Vero spans about $25$ Prony modes and that of A30 about $14$ (\textit{SI Appendix}, Note~\sref{S1} and Table~\sref{S3}). Multi-branch models of other PolyJet materials use comparable numbers of branches~\cite{Mao2015,Wu2016}.

Fig.~\ref{fig:loss}C combines the two master curves into the loss modulus, $E''=E'\tan\delta$, with $E'$ from Eqs.~\eqref{eq:form} to~\eqref{eq:master} and $\tan\delta=\tan[(\Delta/W)\,g_\delta(u)]$. The computed loss modulus follows the measured TTE master curve of the loss modulus across the transition (\textit{SI Appendix}, Fig.~\sref{S2} and Methods). Two master curves, $g(u)$ and $g_\delta(u)$, with the plateaus, $z_c$ and $W$ of each composition thus give the storage modulus, phase angle and loss modulus of the entire family.

\subsection{Prediction of stiffness and damping across composition}\label{sec:tphi}

\begin{figure*}[tbp]
\centering
\includegraphics{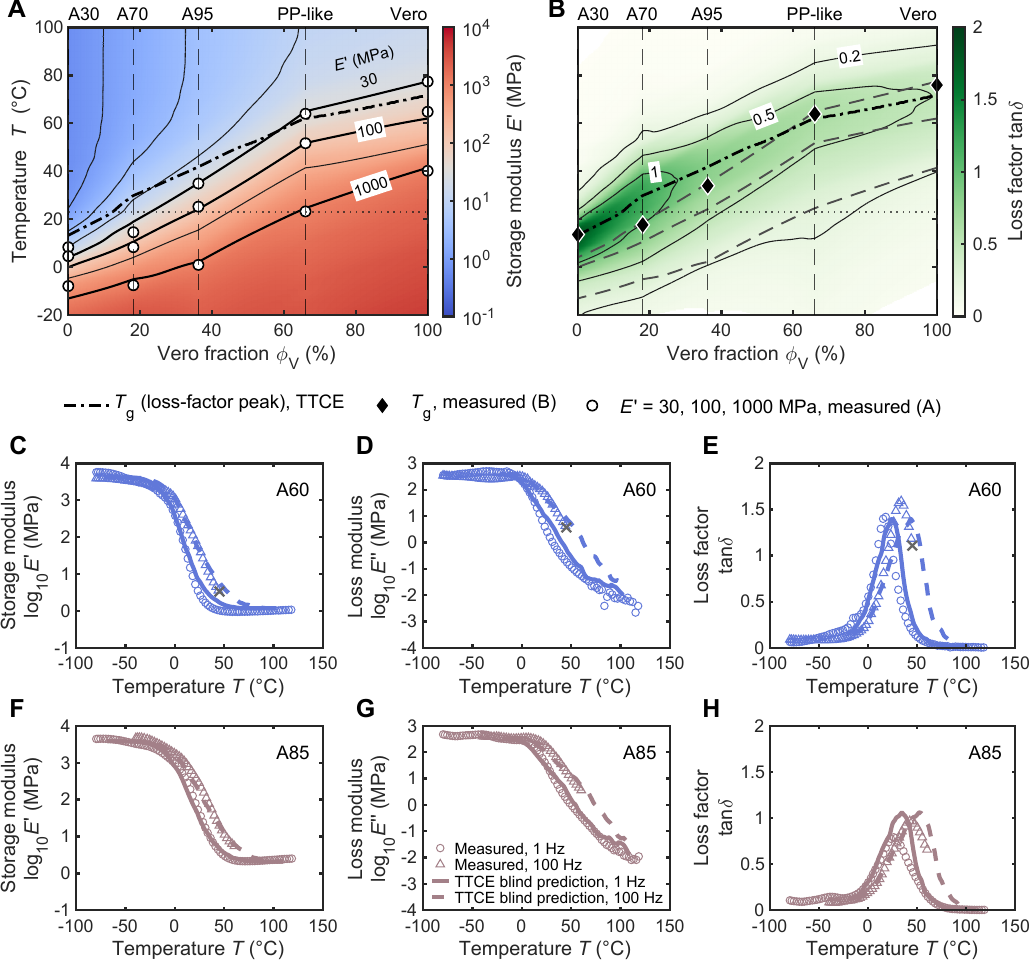}
\caption{\textbf{Prediction of stiffness and damping across composition, and blind tests on two mixtures.} (A) Storage modulus and (B) loss factor at $1$~Hz over Vero fraction and temperature, from the closed form. Position and breadth follow Eqs.~\eqref{eq:shift} and~\eqref{eq:breadth}, and the other parameters are interpolated between the measured compositions (vertical dashed lines). Solid lines are contours of $E'$ ($1$ to $1000$~MPa) in A and of $\tan\delta$ ($0.2$, $0.5$ and $1$) in B. The dash-dotted line is the locus of $T_g$ of the closed form, and the dotted line marks $23~^{\circ}$C. Dashed gray lines in B repeat the $30$, $100$ and $1000$~MPa contours of A. Circles in A mark where the measured $1$~Hz sweeps reach these three moduli (\textit{SI Appendix}, Table~\sref{S4}). Diamonds in B mark the measured $T_g$ (\textit{SI Appendix}, Table~\sref{S1}). (C--H) Blind tests on A60 (C--E) and A85 (F--H), two mixtures not used in the calibration. Storage modulus (C and F), loss modulus (D and G) and loss factor (E and H) against temperature at $1$~Hz (circles) and $100$~Hz (triangles). Lines are the closed form at $1$~Hz (solid) and $100$~Hz (dashed). The cross marks where the A60 specimen fractured at $100$~Hz.}
\label{fig:tphi}
\end{figure*}

Fig.~\ref{fig:tphi} uses the closed form of the TTCE, Eqs.~\eqref{eq:form} to~\eqref{eq:breadth}, to predict the stiffness and damping at every Vero fraction between $0$ and $100\%$. At each Vero fraction, the position $z_c$ and breadth $W$ follow from the linear laws of Eqs.~\eqref{eq:shift} and~\eqref{eq:breadth}. The plateaus, $T_g$ and shift factor are interpolated between the two neighboring measured compositions (Methods). The storage modulus then follows from Eqs.~\eqref{eq:form} to~\eqref{eq:master} and the loss factor from $g_\delta(u)$ of Fig.~\ref{fig:loss}A, with no new parameter.

At $1$~Hz (Fig.~\ref{fig:tphi}A), along the dotted line at $23~^{\circ}$C, the storage modulus increases by three decades of modulus from A30 to Vero. Along a vertical line at fixed composition, the modulus decreases with temperature through the glass transition. The contours of constant modulus therefore run diagonally, nearly parallel to the locus of $T_g$ (dash-dotted line, the peak of the loss factor of Fig.~\ref{fig:tphi}B). At the five measured compositions, the $1$~Hz sweeps reach $30$, $100$ and $1000$~MPa within about $5~^{\circ}$C of the corresponding contours (open circles, \textit{SI Appendix}, Table~\sref{S4}). Along any contour, a change of composition is equivalent to a change of temperature. On the $100$~MPa contour, for example, A70 at $11~^{\circ}$C and PP-like at $52~^{\circ}$C have the same storage modulus. A change of frequency is likewise equivalent to a change of composition, as the measured room-temperature data of Fig.~\ref{fig:gd}A show.

The loss factor of Fig.~\ref{fig:tphi}B follows from the same closed form. Along a vertical line at fixed composition, the loss factor increases with temperature to a peak and then decreases. The peaks of all compositions form a ridge, the locus of $T_g$. From A30 to Vero, the ridge moves to higher temperature, decreases in height and widens, as in the measured sweeps of Fig.~\ref{fig:dma}B. Along a horizontal line at fixed temperature, the loss factor is therefore highest at the composition whose glass transition lies at that temperature. The Agilus-rich compositions damp strongly over a narrow range of temperature, and the Vero-rich compositions damp less over a wider range. The locus of $T_g$ of the closed form lies within about $8~^{\circ}$C of the measured $T_g$ (diamonds) at four of the five compositions. The closed form also agrees with the measured sweeps at $100$~Hz (\textit{SI Appendix}, Fig.~\sref{S4}).

A stricter test is a “blind” prediction whereby the closed form simulates two mixtures, A60 ($\phi_V=13.2\%$) and A85 ($\phi_V=25.3\%$), prior to testing. Fig.~\ref{fig:tphi} C--H compares the predicted and measured sweeps at $1$ and $100$~Hz (the $100$~Hz sweeps end at fracture in A60 and at the noise level in A85). In both mixtures, the predicted storage and loss moduli closely follow the measured responses through the transition (Fig.~\ref{fig:tphi} C, D, F and G). The prediction also captures the shift of the glass transition to higher temperature from $1$ to $100$~Hz. The predicted loss-factor peak shifts with it and lies within several degrees of the measured peak (Fig.~\ref{fig:tphi} E and H). The five calibrated compositions deviate from the master curve $g_\delta(u)$ by a similar amount (Fig.~\ref{fig:loss}A). The closed form thus predicts the stiffness and damping of compositions outside its calibration, at both frequencies and with no adjustment.

\subsection{Temperature--composition equivalence extends to a patterned composite family of different resins}\label{sec:tempcomp}

\begin{figure*}[tbp]
\centering
\includegraphics{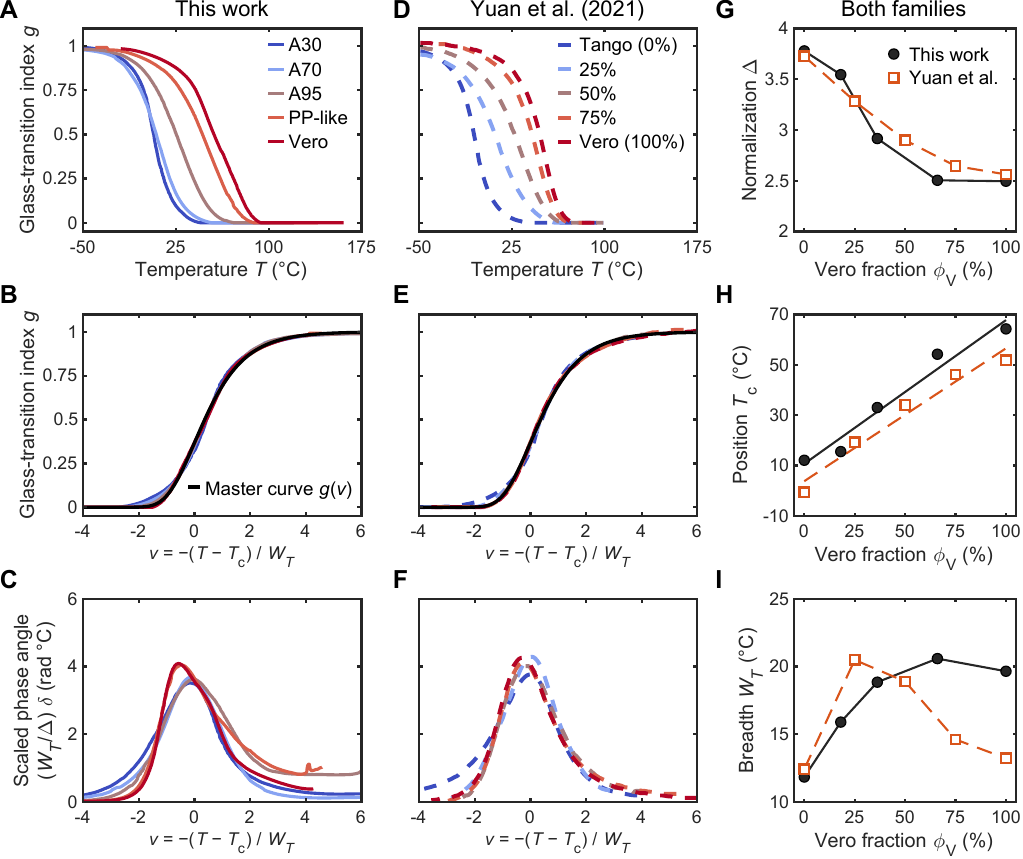}
\caption{\textbf{Temperature--composition equivalence at $1$~Hz, in the present family and in a patterned composite family of different resins.} (A--C) The present family. (D--F) Vero/Tango composites of Yuan et al.~\cite{Yuan2021}, patterned in $300~\mu$m cells and digitized from their published sweeps. (A and D) Glass-transition index $g$ against temperature, normalized by the plateaus of each sweep. (B and E) The same sweeps against $\mathit{v}=-(T-T_c)/W_T$, with the master curve $g(\mathit{v})$ of Eq.~\eqref{eq:master} (black). (C and F) Phase angle multiplied by $W_T/\Delta$, against $\mathit{v}$. (G--I) Normalization $\Delta$, position $T_c$ and breadth $W_T$ against Vero fraction. Lines in H are linear fits, and lines in G and I join the points.}
\label{fig:tempcomp}
\end{figure*}

At a fixed frequency, the reduced variable $u$ of Eq.~\eqref{eq:u} depends only on temperature and composition, and the TTCE reduces to a temperature--composition equivalence. Evaluated at $1$~Hz, the closed form of Section~\ref{sec:fphi} reproduces the measured sweeps of all five compositions (\textit{SI Appendix}, Fig.~\sref{S5}). The plateau normalization, position shift and breadth stretch act along the temperature axis, with a position $T_c$ and a breadth $W_T$ in place of $z_c$ and $W$. The reduced variable becomes $\mathit{v}=-(T-T_c)/W_T$. The normalized sweeps (Fig.~\ref{fig:tempcomp}A) collapse onto the single master curve of Eq.~\eqref{eq:master} (Fig.~\ref{fig:tempcomp}B). The phase angles of the sweeps, scaled by $W_T/\Delta$ as in Fig.~\ref{fig:loss}A, coincide in position and breadth (Fig.~\ref{fig:tempcomp}C). As in Fig.~\ref{fig:gd}, $\Delta$ decreases with the Vero fraction and $T_c$ is linear in it (Fig.~\ref{fig:tempcomp} G and H). The breadth $W_T$ combines $W$ with the temperature dependence of the shift factor (Fig.~\ref{fig:tempcomp}I). At a fixed temperature, the TTCE likewise reduces to a time--composition equivalence (\textit{SI Appendix}, Fig.~\sref{S6}).

Dynamic mechanical data are most often reported as temperature sweeps at a single frequency. The temperature--composition equivalence can therefore be tested on such data, including data on printed systems that differ fundamentally from ours. Yuan et al. studied periodic arrays of $300~\mu$m square cells, each with a stiff Vero inclusion in a soft Tango matrix~\cite{Yuan2021}. This family differs from ours in the printer, the resins, and importantly in the discrete patterning as opposed to voxel level mixing. Yet the same three operations collapse its five compositions, from neat Tango to neat Vero, onto the same master curve (Fig.~\ref{fig:tempcomp} D and E). The scaled phase angles of this family also coincide (Fig.~\ref{fig:tempcomp}F). The normalization $\Delta$ and position $T_c$ follow the same trends with Vero fraction as in the present family (Fig.~\ref{fig:tempcomp} G and H). The breadth $W_T$, in contrast, is larger for every mixture than for either base resin (Fig.~\ref{fig:tempcomp}I). The temperature--composition equivalence thus holds for two fundamentally different printed families, each with parameters of its own, pointing to broad applicability.

\section{Discussion}\label{sec:discussion}

Digitally mixed polymers offer new avenues to design and tailor viscoelasticity. Translation alone, the operation of classical TTE, captures the viscoelasticity of each composition separately but cannot provide an equivalence across the family. The TTCE adds two operations to that translation, a breadth stretch and a plateau normalization. With these three operations, the storage modulus, the phase angle and the relaxation spectrum of all compositions each superpose on one master curve against a single reduced variable. The family therefore shares one distribution of relaxation times. Temperature and frequency, imposed in service, only translate this distribution. Composition, fixed at printing, also stretches it and moves the modulus plateaus.

The physical basis of the TTCE is the mobility of molecular chain segments. An increase in the Vero fraction lengthens the time needed for segmental relaxation, as a decrease in temperature does, and so moves the glass transition to higher temperature. The breadth of the transition reflects the molecular structure of each resin. Vero shows a very broad and asymmetric transition, reflecting a heterogeneous, densely cross-linked molecular structure and hence a wide distribution of relaxation times~\cite{Roland1994,Shundo2021b}. In contrast, Agilus shows a much narrower transition of the same asymmetric shape, reflecting a more uniform molecular structure. Vero also has a rubbery plateau modulus twenty times that of Agilus, owing to a much greater cross-link density. The glassy plateau modulus varies far less, because below the glass transition the stiffness is heavily governed by intermolecular interactions in addition to cross-link density. The three operations of the TTCE therefore have molecular origins. The position shift accounts for the change in segmental mobility, the breadth stretch for the width of the distribution of relaxation times, and the plateau normalization for cross-linking and intermolecular interactions. Each mixture behaves as one material whose molecular mobility is intermediate between those of Agilus and Vero, consistent with its single glass transition, on which the equivalence rests.

The position and breadth of the transition change linearly with the Vero fraction, and the plateaus, $T_g$ and shift factors vary smoothly with it. The closed form therefore gives the response of every composition between those measured, including the intermediate compositions that a database of separately measured materials does not contain~\cite{KarimiMahabadi2026}. Blind predictions for two such compositions agree with their measured stiffness and damping at $1$ and $100$~Hz. For design, each composition enters the closed form through four quantities: its two plateaus, its position and its breadth. Together with the shift factors, these four quantities give stiffness and damping at every temperature and frequency. At a given service temperature and frequency, inverting the closed form gives the composition that reaches a target stiffness, together with its damping. The TTCE thus provides a physical basis for the design of functionally graded and architected multimaterial structures.

Beyond this family, the present work shows how a fabrication variable enters the framework of TTE. Like the present family, the patterned composite family above, made of different resins, shows a single glass transition and obeys the same three operations. In that family, however, the glass transition of each mixture is broader than that of either base resin, consistent with resins that occupy separate regions of each $300~\mu$m cell. In the present family, the breadth of the transition instead changes linearly with the Vero fraction, which suggests that the two base resins are intimately mixed down to the molecular scale. The equivalence therefore extends beyond one printer, one pair of resins and one way of combining them. We expect the TTCE to hold for other 3D printing processes that set composition or degree of cure voxel by voxel and produce a single glass transition. Examples are grayscale and wavelength-selective photopolymerization~\cite{Kuang2019,Yue2023,Kim2025} and multimaterial extrusion that mixes or layers the materials within a filament~\cite{Larson2023,Brown2025}. The present closed form was calibrated on one family and holds for the linear viscoelastic response at small strain. The TTCE can also guide the scaling of finite-strain constitutive models across compositions, as in our model of the same family~\cite{ShenBoyceJMPS}. More broadly, the TTCE makes composition a third variable of viscoelastic behavior, alongside the temperature and frequency that TTE has joined for more than seventy years. A printed polymer family can then be characterized once and designed across its entire range of composition, temperature and frequency.

\matmethods{\subsection*{Materials and printing}
Specimens were printed on a Stratasys PolyJet J826 in High Mix mode, with a nominal layer thickness of $27~\mu$m. The base resins were Agilus~30~Black (A30) and Vero~Ultra~White (Vero), acrylates that are elastomeric and glassy, respectively, at room temperature. Mixing the two at the voxel scale in the manufacturer's prescribed patterns gives the digitally mixed grades A60, A70, A85, A95 and PP-like (RGDA-8530-DM). Support material was removed mechanically, and the specimens were rinsed with water and air-dried, with no post-cure. Our earlier work gives the Vero volume fraction $\phi_V$ of each grade and describes the materials and printing in full~\cite{ShenBoyceJMPS}.

\subsection*{Dynamic mechanical analysis}
The dynamic data of Fig.~\ref{fig:dma} A--C are those of our earlier work~\cite{ShenBoyceJMPS}. Rectangular strips ($25\times5\times1$~mm) were printed flat, with their length along the scan direction of the print head. They were tested in tension on a DMA~850 (TA Instruments) at strain amplitudes of $0.1\%$ or less, within the linear viscoelastic regime. Each run combined the four frequencies with a temperature ramp of $2~^{\circ}$C~min$^{-1}$ from $-80~^{\circ}$C to the rubbery plateau. A60 and A85 (Fig.~\ref{fig:tphi} C--H) were tested on the same instrument, after its servicing, in separate ramps at $1$ and $100$~Hz. At $100$~Hz, the A60 specimen fractured at $47~^{\circ}$C, and the A85 data end at $61~^{\circ}$C, where the signal reached the noise level. The glass transition temperature $T_g$ was taken as the peak of $\tan\delta$ at $1$~Hz (\textit{SI Appendix}, Table~\sref{S1}).

\subsection*{TTE master curves and shift factors}
For each composition, the four frequencies measured at one temperature form an isotherm. The isotherms were shifted along $\log_{10}f$ onto a TTE master curve referenced to $23~^{\circ}$C~\cite{Ferry1980}. The storage modulus, loss modulus and loss factor were aligned jointly, so one shift factor $a_T$ serves all three. Only shifts with sufficient overlap between an isotherm and the assembled curve were accepted. Above $T_g$, where free volume governs segmental motion, the accepted shift factors follow the Williams--Landel--Ferry form~\cite{WLF1955,Doolittle1951}
\begin{equation}
\log_{10}a_T=\log_{10}a_{T_g}-\frac{C_1^{g}\,(T-T_g)}{C_2^{g}+(T-T_g)}.
\label{eq:wlf}
\end{equation}
Below $T_g$, where relaxation is thermally activated, they follow an Arrhenius form~\cite{OConnellMcKenna1999}
\begin{equation}
\log_{10}a_T=\log_{10}a_{T_g}+\frac{E_a}{2.303\,R}\left(\frac{1}{T}-\frac{1}{T_g}\right).
\label{eq:arrhenius}
\end{equation}
Here $C_1^{g}$ and $C_2^{g}$ are constants referenced to $T_g$, $E_a$ is an apparent activation energy, $R$ is the gas constant and $T$ is the absolute temperature. The two branches were fitted jointly and share $\log_{10}a_{T_g}$, so they are continuous at $T_g$ (\textit{SI Appendix}, Table~\sref{S1}). The fits summarize the shift factors, and every calculation in the paper uses the measured shift factors. Each master curve was smoothed along reduced frequency by a running median and a local regression. The smoothed curves, shown in black in Fig.~\ref{fig:dma}C, enter all later steps except the fit of $z_c$ and $W$ below.

\subsection*{Closed form of the storage modulus}
The plateaus $E'_g$ and $E'_r$ were taken as the means over the flat ends of the smoothed TTE master curve of the storage modulus. With the plateaus fixed, $z_c$ and $W$ were fitted by least squares in $\log_{10}E'$ to every shifted point of the master curve (\textit{SI Appendix}, Table~\sref{S2}). The linear laws of Eqs.~\eqref{eq:shift} and~\eqref{eq:breadth} were then fitted to the five values of $z_c$ and of $W$. Like $z$, $z_c$ is referenced to $23~^{\circ}$C. A different reference temperature shifts $z$ and $z_c$ equally and leaves $W$ and $u$ unchanged. The collapse of Fig.~\ref{fig:gd}F is measured by the root-mean-square deviation of the index from $g(u)$ over all points of the five TTE master curves.

\subsection*{Loss and relaxation spectrum}
The scaling of the loss follows from the first approximation of Schwarzl and Staverman~\cite{SchwarzlStaverman1953}. For a Prony series of relaxation modes~\cite{Ferry1980}, the loss modulus and the slope $dE'/d\ln\omega$ are both sums over the modes, with weights about one decade of relaxation time wide. Where the relaxation spectrum varies little over that width, $\tan\delta\approx(\pi/2)\,d\log_{10}E'/d\log_{10}\omega$. With Eq.~\eqref{eq:form}, $d\log_{10}E'/d\log_{10}\omega$ is $\Delta/W$ times a function of $u$ alone, so $(W/\Delta)\tan\delta$ is approximately a function of $u$ alone. The same scaling is applied to the phase angle, which approaches $\tan\delta$ where the loss is small. The approximation fixes the factor $W/\Delta$, and the shape of $g_\delta(u)$ is taken from the measurements, as the mean of the five scaled curves of Fig.~\ref{fig:loss}A.

The Prony series of Fig.~\ref{fig:loss}B was fitted jointly to the TTE master curves of the storage modulus and loss factor of each composition. The mode strengths were kept positive, and a penalty kept the spectrum smooth (\textit{SI Appendix}, Note~\sref{S1}). The computed loss modulus of Fig.~\ref{fig:loss}C agrees with the measured TTE master curve to within $0.1$ to $0.3$ decade of modulus in root mean square.

\subsection*{Prediction across composition}
Fig.~\ref{fig:tphi} is computed as described in Section~\ref{sec:tphi}, and \textit{SI Appendix}, Fig.~\sref{S4} is computed in the same way at $100$~Hz. The plateaus and $T_g$ are interpolated linearly in $\phi_V$, and the shift factor is interpolated at a fixed distance $T-T_g$ from the glass transition. The locus of $T_g$ is the peak of the computed loss factor. The circles of Fig.~\ref{fig:tphi}A are read from the measured $1$~Hz sweeps at the same levels (\textit{SI Appendix}, Note~\sref{S2}). The plateaus, $T_g$ and shift factor of A60 and A85 (Fig.~\ref{fig:tphi} C--H) are interpolated in the same way, between their two neighboring measured compositions.

\subsection*{Temperature--composition equivalence}
For each $1$~Hz sweep, the position $T_c$, the breadth $W_T$ and the glassy plateau were fitted by least squares in $\log_{10}E'$. The rubbery plateau was taken as the mean over the last $10~^{\circ}$C of the sweep. Both families were treated the same way. The sweeps of Yuan et al., measured in tension at $1$~Hz on cooling, were digitized from their published figure after calibration on the axis ticks. The article is open access under a CC~BY~4.0 license.

\subsection*{Use of AI tools}
Claude Opus 5.5 (Anthropic) was used for final language editing and to assist with the data-analysis scripts. The authors verified all content and take full responsibility for the paper.
}

\showmatmethods{}

\dataavail{Study data are included in the article and/or \textit{SI Appendix}. Fig.~\ref{fig:tempcomp} also uses previously published data~\cite{Yuan2021}.}

\acknow{This work was supported by Columbia University. The authors thank Prof.\ Yevgeniy Yesilevskiy for laboratory management and equipment maintenance, and Dr.\ Will Hunnicutt for DMA training and for coordinating equipment use at the Carleton Laboratory, Columbia University.}

\showacknow{}

\subsection*{Author contributions}
B.S.: conceptualization, methodology, investigation, formal analysis, data curation, software, validation, visualization, writing (original draft, review and editing). S.K.K.: methodology, writing (review and editing). M.C.B.: conceptualization, methodology, supervision, funding acquisition, resources, writing (review and editing).

\subsection*{Competing interests}
The authors declare no competing interests.

\bibliography{tcs_refs}

\clearpage
\onecolumn


\section*{Supplementary material}

# Supporting Information for

## Time–temperature–composition equivalence of the glass transition in digitally mixed 3D-printed polymers

Beijun Shen, Sanat K. Kumar and Mary C. Boyce

Corresponding author: Mary C. Boyce. E-mail: mb3814@columbia.edu

**This PDF file includes:**

Supporting text
Figs. S1 to S6
Tables S1 to S4
SI References

*Figs. 1–6, Eqs. 1–7, Sections 1.1–1.5 and Methods refer to the main text. S-numbered figures, tables and notes are in this document.*

## Transition temperatures

The onset of the $E'$ downturn (Fig. S1A) is the intersection of the extrapolated glassy plateau with the tangent at the steepest decrease of $\log_{10} E'$.

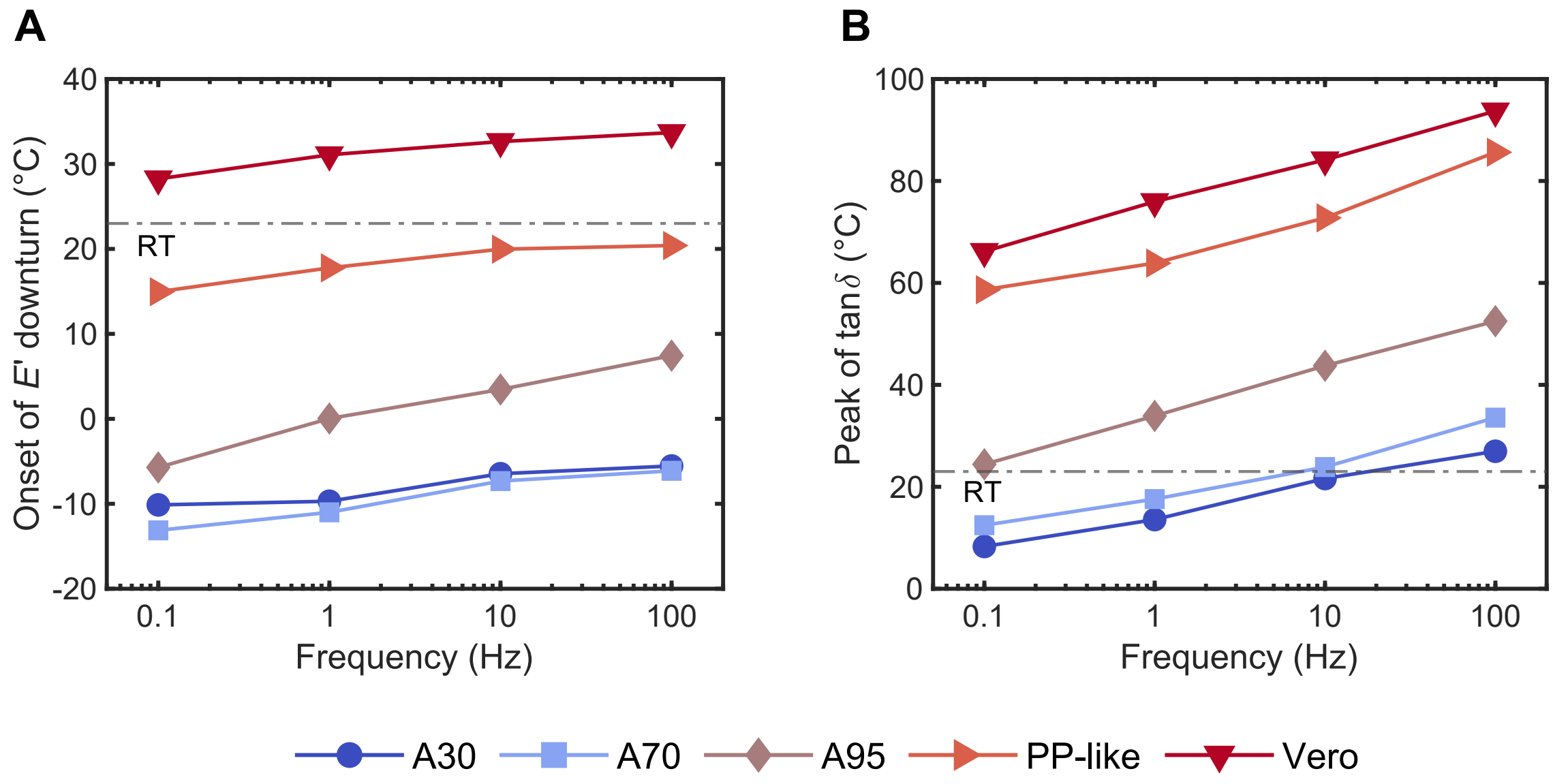


**Fig. S1: Transition temperature against frequency.** (A) From the onset of the $E'$ downturn and (B) from the $\tan\delta$ peak, for the five compositions at 0.1, 1, 10 and 100 Hz. The temperatures are those of the sweeps of Fig. 2 A and B. The onset increases by about 2 to 4 °C and the peak by about 6 to 9 °C per decade of frequency. The dash-dotted line marks room temperature, 23 °C (RT). The sweeps were first reported in our earlier work [1].

## TTE master curves and parameters of the closed form

Table S1 lists $T_g$ and the Arrhenius and Williams–Landel–Ferry fits to the shift factors of the five compositions (Fig. 2D). Fig. S2 shows the TTE master curves of each composition (Fig. 2C). Table S2 lists the parameters of the closed form fitted to them, used in Figs. 3–5 and Figs. S4 to S6.

**Table S1: Glass transition temperature and shift-factor fits of the five compositions.** $T_g$ is the temperature of the $\tan\delta$ peak at 1 Hz. $E_a$ is the apparent activation energy of Eq. (7) below $T_g$. $C_1^g$ and $C_2^g$ are the constants of Eq. (6) above it, referenced to $T_g$. $C_1^g/C_2^g$ is the magnitude of the slope of $\log_{10} a_T$ just above $T_g$, in decades of $a_T$ per degree Celsius. An asterisk marks a value at the upper bound of the fit. For these two compositions, the measured window above $T_g$ is short compared with $C_2^g$, so only the ratio of the two constants is determined.

| Composition | $T_g$ (°C) | $E_a$ (kJ mol$^{-1}$) | $C_1^g$ | $C_2^g$ (°C) | $C_1^g/C_2^g$ (°C$^{-1}$) |
|---|---|---|---|---|---|
| A30 | 13.6 | 350 | 32.4 | 245 | 0.13 |
| A70 | 17.6 | 366 | 47.4 | 400* | 0.12 |
| A95 | 33.9 | 277 | 61.3 | 400* | 0.15 |
| PP-like | 63.9 | 168 | 11.4 | 28.9 | 0.39 |
| Vero | 75.9 | 410 | 24.4 | 190 | 0.13 |

**Table S2: Parameters of the closed form for the storage modulus.** Glassy and rubbery plateau values of the storage modulus $E'_g$ and $E'_r$, normalization $\Delta = \log_{10} E'_g - \log_{10} E'_r$, position $z_c$ and breadth $W$ of Eqs. (1) and (2). The plateaus are read from the TTE master curve of the storage modulus of each composition, referenced to 23 °C, and $z_c$ and $W$ are fitted to it (Methods). The values are plotted in Fig. 3 C–E, with $\Delta$ in decades of modulus and $z_c$ and $W$ in decades of reduced frequency. The column $\Delta/W$ is the ratio that scales the phase angle in Fig. 4A. The last column is the stretching exponent $\beta = 1/(W \ln 10)$ of Eq. (3) written in reduced frequency (Section 1.2).

| Composition | $\phi_V$ (%) | $E'_g$ (MPa) | $E'_r$ (MPa) | $\Delta$ | $z_c$ | $W$ | $\Delta/W$ | $\beta$ |
|---|---|---|---|---|---|---|---|---|
| A30 | 0 | 3550 | 0.64 | 3.75 | 2.10 | 2.89 | 1.30 | 0.150 |
| A70 | 18.1 | 4430 | 1.48 | 3.47 | 1.37 | 3.01 | 1.15 | 0.144 |
| A95 | 36.2 | 3180 | 3.57 | 2.95 | −1.48 | 3.49 | 0.85 | 0.124 |
| PP-like | 66 | 5490 | 9.70 | 2.75 | −4.85 | 3.93 | 0.70 | 0.111 |
| Vero | 100 | 5090 | 12.35 | 2.61 | −8.79 | 4.68 | 0.56 | 0.093 |

# Note S1: Prony series

This note checks the Prony series behind the relaxation spectra of Fig. 4B (Methods). With one mode per decade of relaxation time, the series closely reproduces the smoothed TTE master curves of the storage modulus and loss factor (Fig. S3 A and B, dashed). Fig. S3C shows the resulting spectra. Fig. 4B scales them with the position, breadth and plateaus of each composition.

Fig. S3 D–L test the spacing of the modes for A30, A95 and Vero, the two resins and an intermediate mixture. At half a mode per decade of relaxation time, the reconstructed loss factor ($\tan\delta$) of each composition oscillates about the smoothed TTE master curve and overshoots its peak. From one mode per decade of relaxation time upward, the reconstructed curves and spectra no longer change for any composition. One mode per decade of relaxation time, as used in the main text, is therefore sufficient across the family.

The number of modes, however, increases with the span of the transition. At one mode per decade of relaxation time, 14 modes lie within the transition of A30, which spans 13 decades of reduced frequency. The transition of Vero spans 24 decades of reduced frequency and contains 25 modes (Table S3). Vero therefore needs nearly twice as many modes as A30. Published multi-branch models of other PolyJet materials show the same trend. Wu et al. [2] used 27 branches for each of two digital materials with glass transitions above room temperature, DM8530 and DM9895, and 15 for the elastomeric TangoBlack+. Mao et al. [3] used 16 branches for seven digital materials of VeroWhite and TangoBlack and noted that the number of branches depends on the width of the glass transition.

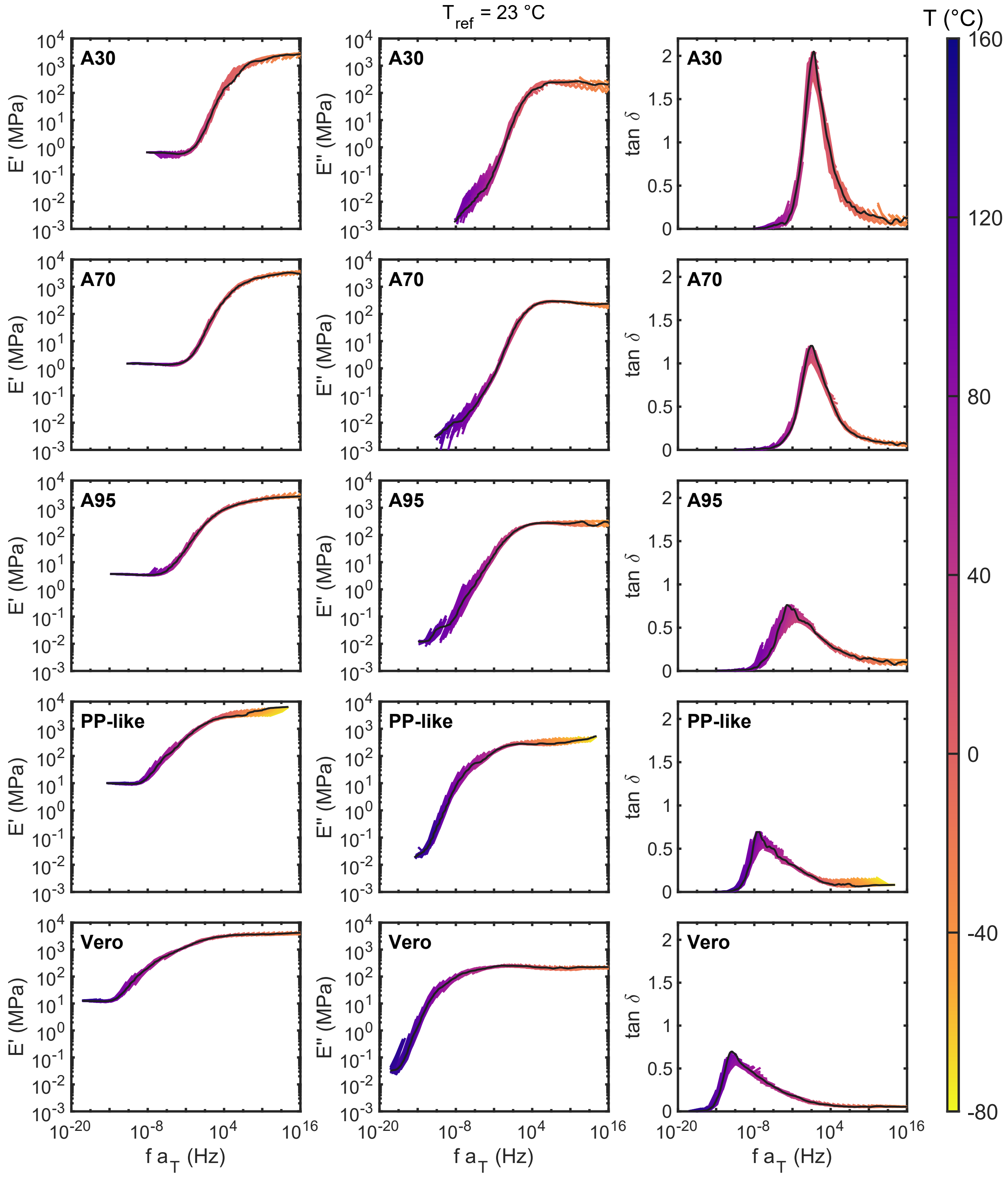


**Fig. S2: TTE master curves of the five compositions.** Storage modulus (left), loss modulus (middle) and loss factor (right) against reduced frequency $f\,a_T$, referenced to 23 °C, one row per composition. Each isotherm is shifted by one shift factor $a_T$ for all three quantities and colored by its temperature. The black curves are the smoothed TTE master curves used in the main text (Methods).

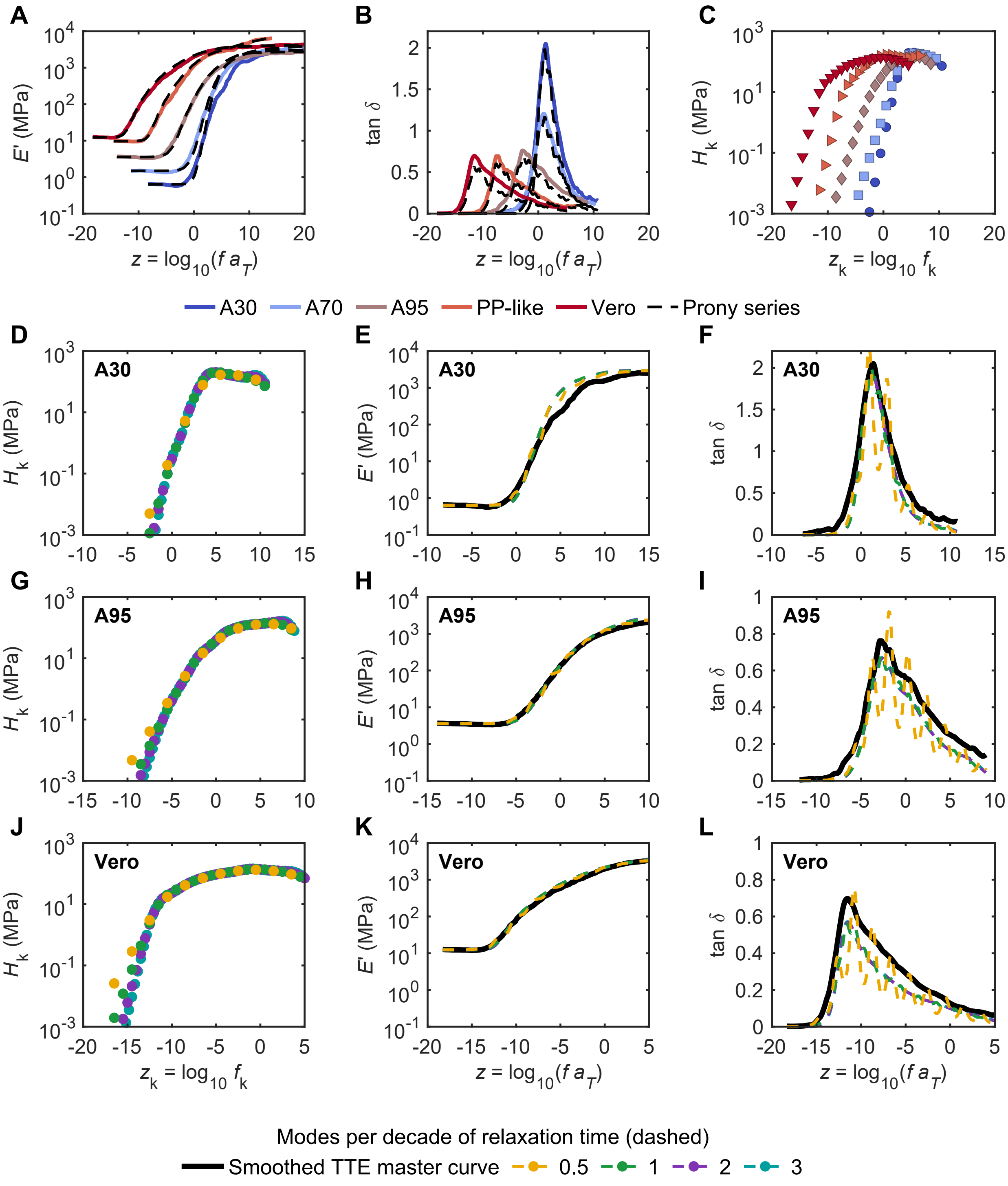


**Fig. S3: Prony series and the spacing of the modes.** (A) Storage modulus and (B) loss factor of the five compositions against reduced frequency $z$, referenced to 23 °C. Colored curves are the smoothed TTE master curves and dashed curves are those reconstructed from the Prony series with one mode per decade of relaxation time. (C) Relaxation strength $H_k$ per unit $\ln \tau$ of each mode against the reduced frequency $z_k$ at which it relaxes, symbols as in Fig. 4. Modes weaker than $10^{-3}$ MPa, the lower limit of the axis, lie on the rubbery side of each transition and together contribute less than 0.2% of the rubbery modulus. (D–L) Spectra (left), storage modulus (middle) and loss factor (right) of A30 (D–F), A95 (G–I) and Vero (J–L). The series have 0.5, 1, 2 and 3 modes per decade of relaxation time. Thick black curves are the smoothed TTE master curves. The curves reconstructed at one mode per decade of relaxation time and above coincide.

Yuan et al. [4] described VeroWhite with 27 branches spanning 17 decades of relaxation time. Ge et al. [5] spaced the relaxation times of the branches one decade of relaxation time apart, as here.

**Table S3: Transition span and number of Prony modes.** The span of the transition of each composition runs from 5 to 95% of the increase of $\log_{10} E'$ between its plateaus, in decades of reduced frequency. The modes in the span are counted at one mode per decade of relaxation time.

| Composition | Span (decades of reduced frequency) | Modes in span |
|---|---|---|
| A30 | 13.2 | 14 |
| A70 | 14.1 | 15 |
| A95 | 16.3 | 16 |
| PP-like | 16.3 | 16 |
| Vero | 24.3 | 25 |

## Prediction of stiffness and damping at 100 Hz

Fig. S4 repeats Fig. 5 at 100 Hz, with the same closed form and no new parameter (Section 1.4).

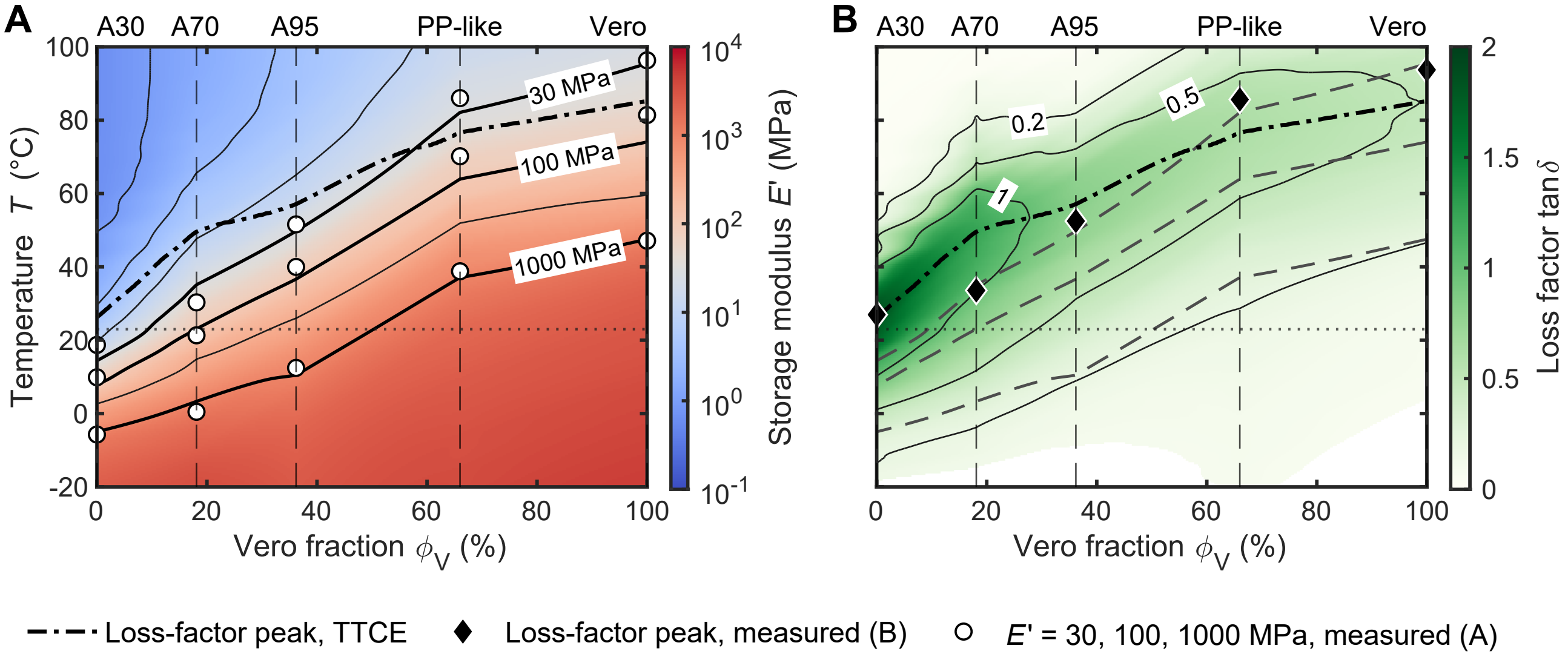


**Fig. S4: Prediction of stiffness and damping at 100 Hz.** (A) Storage modulus and (B) loss factor at 100 Hz over Vero fraction and temperature, from the closed form, as in Fig. 5. Solid lines are contours of $E'$ (1 to 1000 MPa) in A and of $\tan\delta$ (0.2, 0.5 and 1) in B. The dash-dotted line is the locus of the loss-factor peak of the closed form, and the dotted line marks 23 °C. Dashed gray lines in B repeat the 30, 100 and 1000 MPa contours of A. Circles in A mark where the measured 100 Hz sweeps reach these three moduli, within 7.3 °C of the contours. Diamonds in B mark the measured loss-factor peaks at 100 Hz (Fig. S1B), within 9 °C of the locus at four of the five compositions.

## Note S2: Measured values in Fig. 5

The computed and measured temperatures of Table S4 differ by at most 5.2 °C.

**Table S4: Temperatures at which the storage modulus reaches 1000, 100 and 30 MPa at 1 Hz.** Computed values are from the closed form (Fig. 5A), and measured values from the 1 Hz sweeps (Methods). All values are in °C.

| | 1000 MPa | | 100 MPa | | 30 MPa | |
|---|---|---|---|---|---|---|
| Composition | computed | measured | computed | measured | computed | measured |
| A30 | −13.1 | −7.9 | −0.1 | 4.6 | 4.1 | 8.3 |
| A70 | −5.1 | −7.6 | 10.5 | 8.3 | 18.9 | 14.5 |
| A95 | 2.1 | 0.9 | 24.2 | 25.2 | 35.0 | 34.9 |
| PP-like | 24.6 | 23.2 | 51.8 | 51.7 | 65.0 | 64.0 |
| Vero | 41.4 | 40.1 | 62.0 | 64.8 | 77.4 | 77.4 |

# Temperature–composition equivalence at 1 Hz

Fig. S5 compares the measured storage modulus at 1 Hz with the closed form (Section 1.5).

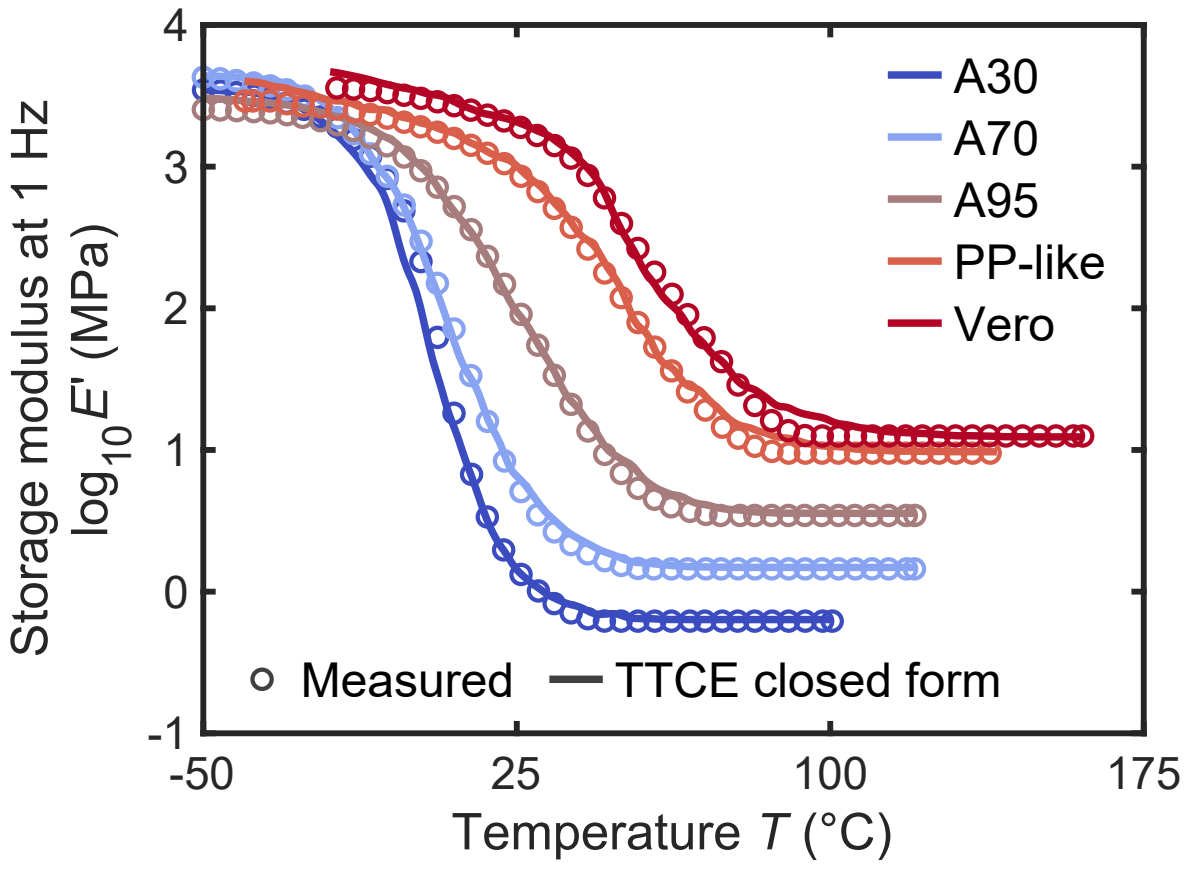


**Fig. S5: Storage modulus at 1 Hz from the closed form.** Symbols are the measured sweeps of the five compositions. Lines are computed from Eqs. (1) to (3) with the parameters of Table S2 and the measured shift factors, with no further parameter.

# Time–composition equivalence at 23 °C

Fig. S6 shows the time–composition equivalence at 23 °C, in the form of Fig. 6 A–C (Section 1.5).

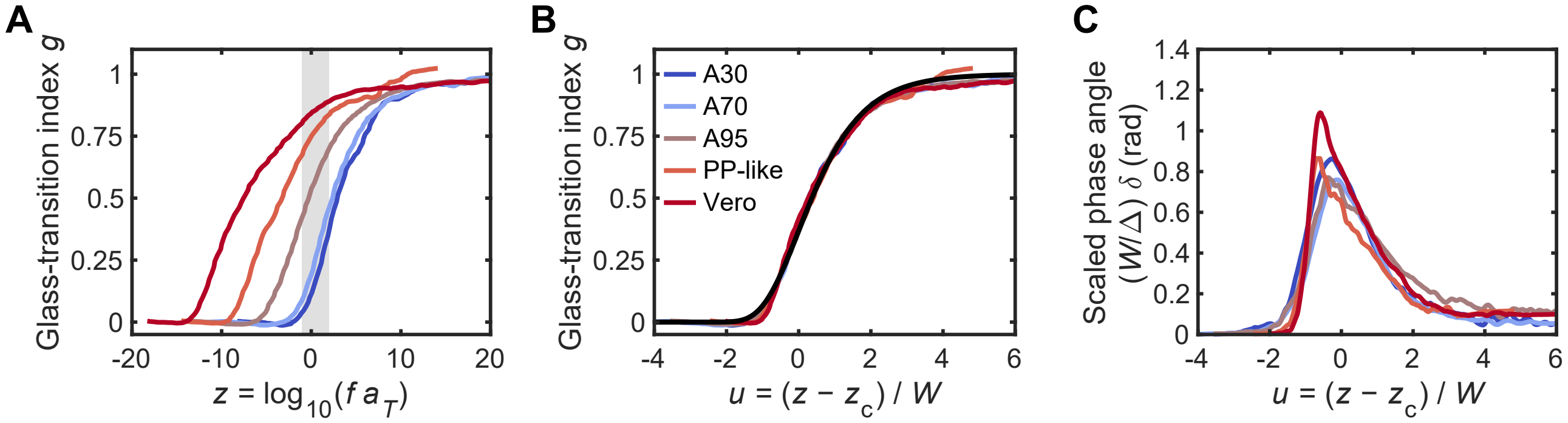


**Fig. S6: Time–composition equivalence at 23 °C.** (A) Glass-transition index $g$ (Eq. (1)) of each composition against reduced frequency $z$ at 23 °C, from the smoothed TTE master curves. The gray band spans the four measured frequencies, 0.1, 1, 10 and 100 Hz. (B) The same curves against $u$ (Eq. (2)), with $g(u)$ of Eq. (3) (black). (C) Phase angle multiplied by $W/\Delta$, against $u$. Parameters are those of Table S2. B and C show the curves of Figs. 3F and 4A in this form.

# SI References



\end{document}